\documentstyle[12pt]{article}

\begin{document}

\vskip 1cm

\begin{center}
THE THREE-DIMENSIONAL BTZ BLACK HOLE AS A CYLINDRICAL SYSTEM
IN FOUR-DIMENSIONAL GENERAL RELATIVITY \\
\vskip 1cm
{\bf Jos\'e P. S. Lemos} \\
\vskip 0.3cm
{\scriptsize  Departamento de Astrof\'{\i}sica,
	      Observat\' orio Nacional-CNPq,} \\
{\scriptsize  Rua General Jos\'e Cristino 77,
	      20921 Rio de Janeiro, Brasil,} \\
{\scriptsize  \&} \\
{\scriptsize  Departamento de F\'{\i}sica,
	      Instituto Superior T\'ecnico,} \\
{\scriptsize  Av. Rovisco Pais 1, 1096 Lisboa, Portugal,} \\
\vskip 0.6cm
{\bf Vilson T. Zanchin} \\
\vskip 0.3cm
{\scriptsize  Departamento de F\'{\i}sica-CCNE,
	      Universidade Federal de Santa Maria} \\
{\scriptsize  97119-900 Santa Maria, RS, Brazil.} \\
\end{center}

\bigskip

\begin{abstract}
\noindent
It is shown how to transform the three dimensional BTZ black hole into a
four dimensional cylindrical black hole (i.e., black string)
in general relativity. This process is identical to the transformation
of a point particle in three dimensions into a straight cosmic string
in four dimensions.
\\

\end{abstract}

\newpage

\noindent
{\bf 1. Introduction}

\vskip 3mm

Black holes were predicted within four dimensional general
relati\-vi\-ty as objects emerging from complete gravitational
collapse of massive objects, such as stars \cite{whee}.
They have also appeared as exact solutions of
se\-ve\-ral gravity theories in two, three, four and higher dimensions.
String theory  has provided a great variety
of black hole solutions as well as extended black objects such as
black  strings and black membranes \cite{horo}.
The study of black holes in dimensions lower than four has proved
fruitful for a better understanding of some physical features
in a black hole geometry.

Recently, it has been shown \cite{bana} that three
dimensional (3D) general relativity with a negative cosmological
constant admits a black hole solution with constant curvature, the
BTZ black hole.
This black hole has mass and angular momentum.
It has  many features similar to the Kerr black holes
in four dimensional (4D) general relativity.
It is interesting, as well as important, to formulate other theories
for the same solution. For instance, the BTZ is also a solution
to string theory in three dimensions \cite{welc}.
Our aim in this paper is to show that the BTZ black hole is also a
black hole solution of cylindrical general relativity.

\vskip 1cm

\noindent
{\bf 2. From 3D to 4D General Relativity}

\vskip 3mm
In order to set the nomenclature we write the  action in $D$ dimensions
as
\begin{equation}
S^{(D)} = {S^{(D)}}_{\rm g} + {S^{(D)}}_{\rm matter},
                         \label{eq:1}
\end{equation}
where $D=3,4$.
${S^{(D)}}_{\rm g}$ is the gravitational Einstein-Hilbert action ($G=c=1$),
\begin{equation}
{S^{(D)}}_{\rm g} = \frac{1}{16\pi }\int d^Dx \sqrt{-g^{(D)}}
(R^{(D)} - 2\Lambda),
                   \label{eq:2}
\end{equation}
where ${g^{(D)}}$ is the determinant of the metric,
$R^{(D)}$ is the  Ricci scalar and $\Lambda$ is the cosmological
constant. The matter action can be written as
\begin{equation}
{S^{(D)}}_{\rm matter}= \int d^Dx \sqrt{-g^{(D)}}
{{\cal L}^{(D)}}_{\rm matter},
                        \label{eq:3}
\end{equation}
where ${{\cal L}^{(D)}}_{\rm matter}$ is the Lagrangian for the matter.
Variation of (\ref{eq:1}) with respect to ${g^{(D)}}_{ab}$ yields the
equations of motion,
\begin{equation}
{G^{(D)}}_{ab} + \Lambda {g^{(D)}}_{ab} = 8\pi  {T^{(D)}}_{ab},
                        \label{eq:4}
\end{equation}
where ${G^{(D)}}_{ab}$ is the Einstein tensor and
${T^{(D)}}_{ab}$ is the energy-momentum tensor, defined by
${T^{(D)}}^{ab} = \frac{2}{\sqrt{-g^{(D)}}}
\frac{\delta (\sqrt{-g^{(D)}}
{{\cal L}^{(D)}}_{\rm matter})}{\delta {g^{(D)}}_{ab}}$.

Now, in 3D we write the metric in the form,
\begin{equation}
d{s^{(3)}}^2 = {g^{(3)}}_{ab} dx^a dx^b.
                        \label{eq:5}
\end{equation}
In addition, in 3D the Riemann tensor can be written as
\begin{equation}
{{R^{(3)}}^{ab}}_{cd}=\epsilon^{abe}\epsilon_{cdf}  {{G^{(3)}}^f}_e.
                        \label{eq:6}
\end{equation}
Thus, from equation (\ref{eq:4}),
for a source free region (${T^{(3)}}_{ab}=0$) one has a space of
constant curvature. If one further sets $\Lambda=0$, one has flat spacetime.

Now, we want to relate general relativity in 3D with general relativity
in 4D, in such a way that a solution in 3D is also a solution in 4D.
This can be achieved by setting the 4D metric as,
\begin{equation}
d{s^{(3)}}^2 = {g^{(3)}}_{ab} dx^a dx^b + dz^2,
                        \label{eq:7}
\end{equation}
where $z$ is a Killing direction in the 4D spacetime. Then, the
determinants of the metric are the same $g^{(4)}=g^{(3)}$,
the Ricci scalars are also the same,
$R^{(4)} = R^{(3)}$ and through dimensional reduction one obtains
3D from 4D general relativity.
Thus, having found a 3D solution one can now find the corresponding 4D
solution by choosing an apropriated energy-momentum tensor in 4D.
In the next section we will do this for the well known example of
3D point particles and 4D cosmic strings. Afterwards we will find the
4D counterpart of the 3D BTZ black hole.
\vskip 1cm

\noindent
{\bf 3. The example of 3D point particles and 4D cosmic strings}

\vskip 3mm

For a point particle with mass $m$ and spin $s$ in 3D
(with cartesian coordinates ($t,x,y$))
the energy-momentum tensor can be written as,
\begin{equation}
{{T^{(3)}}^0}_0  = m \delta(x)\delta(y),
                     \label{eq:8}
\end{equation}
\begin{equation}
{{T^{(3)}}^i}_0  = s \epsilon^{ij}\partial_j\delta(x)\delta(y),
                        \label{eq:9}
\end{equation}
\begin{equation}
{{T^{(3)}}^i}_j=0,
                       \label{eq:10}
\end{equation}
where $i,j=x,y$. Then by integrating Einstein's equations
(with $\Lambda=0$) one finds the following metric \cite{jackiw1},
\begin{eqnarray}
&{ds^{(3)}}^2 = -dt^2 -8sdtd\varphi +dr^2 +r^2(1-4m)^2 d\varphi^2, &
\nonumber\\
& -\infty< t < \infty,\quad 0\leq r<\infty, \quad0\leq\varphi<2\pi, &
                        \label{eq:11}
\end{eqnarray}
where ($r,\varphi$) are the polar coordinates associated to ($x,y$).
There are two Killing vectors,
$\frac{\partial}{\partial t}$ and
$\frac{\partial}{\partial\varphi}$, and
the full symmetry group of this spacetime is $R\times SO(2)$ \cite{brown1}.
By the cordinate transformation $\overline{t} = t - 4s\varphi$ and
$\overline\varphi = (1-4m)\varphi$ one obtains a flat spacetime,
\begin{equation}
{ds^{(3)}}^2 = - d\overline{t}^2 + dr^2 + r^2 d\overline\varphi^2,
                        \label{eq:12}
\end{equation}
but now with the obligatory identifications
$(\overline{t}, r, \overline{\varphi}) = (\overline{t} + 8\pi s, r,
\overline\varphi
+ 2\pi(1-4m))$, which mean space is conical and time is helical.

Now, from equations (\ref{eq:5}) and (\ref{eq:7})
we can put metric (\ref{eq:11}) in the 4D form,
\begin{equation}
{ds^{(3)}}^2 = -dt^2 -8sdtd\varphi +dr^2
+r^2(1-4m)^2 d\varphi^2+dz^2,
                      \label{eq:13}
\end{equation}
with $-\infty<z<\infty$. In 4D, Einstein's equations for metric
(\ref{eq:13}) can be split into
\begin{equation}
{G^{(3)}}_{ab} = 8\pi  {T^{(3)}}_{ab}
                      \label{eq:14}
\end{equation}
\begin{equation}
{G^{(3)}}_{zz} = 8\pi  {T^{(3)}}_{zz}
                     \label{eq:15}
\end{equation}
and the cross equation ${G^{(3)}}_{az}$ is identically satisfied.
Then, from equation (\ref{eq:4}) we can deduce,
\begin{equation}
G^{(4)} = - R^{(4)} = -R^{(3)} = 8\pi T^{(4)},
                        \label{eq:16}
\end{equation}
where $G^{(4)}$ is the trace of the Einstein tensor in 4D, etc.
{}From (\ref{eq:14}) and (\ref{eq:8})--(\ref{eq:10}) we also deduce
\begin{equation}
R^{(3)} = -16\pi  T^{(3)} = -16\pi  {{T^{(3)}}^0}_0.
                        \label{eq:17}
\end{equation}
But $T^{(4)} =  T^{(3)} +  {{T^{(3)}}^z}_z = {{T^{(3)}}^0}_0
+ {{T^{(3)}}^z}_z$. Then from  (\ref{eq:16}) and  (\ref{eq:17})
and the fact that ${{T^{(4)}}^0}_0={{T^{(3)}}^0}_0$ we
obtain,
\begin{equation}
{{T^{(4)}}^z}_z = {{T^{(4)}}^0}_0.
                        \label{eq:18}
\end{equation}
Therefore, metric (\ref{eq:13}) is a solution of 4D general relativity if
the energy-momentum tensor is given by,
\begin{equation}
{{T^{(4)}}^0}_0  = m \delta(x)\delta(y),
                     \label{eq:19}
\end{equation}
\begin{equation}
{{T^{(4)}}^i}_0  = s \epsilon^{ij}\partial_j\delta(x)\delta(y),
                        \label{eq:20}
\end{equation}
\begin{equation}
{{T^{(4)}}^z}_z  = m \delta(x)\delta(y),
                     \label{eq:21}
\end{equation}
with $i,j=x,y$ and the other components are zero.
This is the energy-momentum tensor of a straight
cosmic string \cite{vilenkin1}. But now $m$ and $s$ are the mass
per unit length and the angular momentum per unit length of the string,
respectively. There are three Killing vectors,
$\frac{\partial}{\partial t}$,
$\frac{\partial}{\partial\varphi}$ and
$\frac{\partial}{\partial z}$, and
the full symmetry group of this spacetime is $R^2\times SO(2)$.

\vskip 1cm

\noindent
{\bf 4. The 3D BTZ Black Hole and the Black String in 4D}

\vskip 3mm

The 3D BTZ black hole appears as a solution of Einstein's gravity
with a negative cosmological term, $\alpha^2\equiv-\Lambda>0$,
and ${T^{(3)}}_{ab}=0$. The metric
is given by \cite{bana,bhtz},
\begin{equation}
{ds^{(3)}}^2 = - (\alpha^2 r^2 - 8M)dt^2 -8J dt d\varphi +
\frac{dr^2}{\alpha^2 r^2 - 8M + \frac{16J^2}{r^2}} + r^2d\varphi^2,
                        \label{eq:22}
\end{equation}
where $M$ and $J$ are the mass and angular momentum of the black hole.
(Note that the normalization for $M$ and $J$ is different from \cite{bana}.
We are using $G=1$, instead of $G=\frac18$).
This solution has constant curvature and appears through identification
of points in the anti-de Sitter spacetime.
The asymptotic symmetry group is the conformal group in two dimensions
wich has the anti-de Sitter group $SO(2,2)$ as a subgroup
\cite{brown1,bhtz}.

We note that the cosmological
term in Einstein's equations (i.e., $\Lambda {g^{(3)}}_{ab}$ in
equation (\ref{eq:4}))  can be regarded, if one wishes,
as an energy-momentum tensor for a perfect fluid
\cite{macrea}. Indeed, if we write
\begin{equation}
{T^{(3)}}_{ab} = (\rho + p)u_au_b + pg_{ab},
                        \label{eq:23}
\end{equation}
where $\rho$ is the energy-density and $p$ is the pressure,
and set
\begin{equation}
8\pi \rho=-8\pi p=-\alpha^2,
                        \label{eq:23b}
\end{equation}
we obtain the
cosmological term.
In this description the cosmological term does not appear explicitly
in Einstein's equations (\ref{eq:4}), which are then given by,
${G^{(3)}}_{ab} = 8\pi  {T^{(3)}}_{ab}$, with
${T^{(3)}}_{ab}$ given by equations (\ref{eq:23}) and (\ref{eq:23b}).
This fluid obeys the strong but not the weak
energy condition since its energy-density is negative \cite{hawkingellis}.

Now we want to translate the above black hole solution  (\ref{eq:22})
into a solution of 4D general relativity. Then using  (\ref{eq:7})
we have
\begin{equation}
{ds^{(4)}}^2 = - (\alpha^2 r^2 - 8M)dt^2 -8J dt d\varphi +
\frac{dr^2}{\alpha^2 r^2 - 8M + \frac{16J^2}{r^2}} + r^2d\varphi^2
 +dz^2.
                             \label{eq:24}
\end{equation}
Using the formalism of Brown and York \cite{brown2} we find that $M$ and
$J$ are now the mass per unit length and the angular momentum per
unit length, respectively, of the black string solution
given in (\ref{eq:24})
(for an analysis of a more general case see \cite{lemoszanchin}).

To know what is the stress-energy tensor for this solution we take the trace
of  (\ref{eq:4}) to find
\begin{equation}
-R^{(4)}+4\Lambda = 8\pi  T^{(4)}.
                       \label{eq:25}
\end{equation}
But,
\begin{equation}
R^{(4)} = R^{(3)} = 6\Lambda = -6\alpha^2.
                       \label{eq:26}
\end{equation}
Then from  (\ref{eq:25}) and  (\ref{eq:26}) we obtain,
\begin{equation}
8\pi  T^{(4)} = -2\Lambda = 2\alpha^2.
                       \label{eq:27}
\end{equation}
Since here ${T^{(3)}}_{ab}=0$, we are left with
\begin{equation}
8\pi {{T^{(4)}}^z}_{z}= 2\alpha^2,
                        \label{eq:28}
\end{equation}
and the other components are equal to zero. One can see this fluid as
one having pressure in the $z$ direction alone, and no
energy-density. Such a fluid also obeys the strong but not the weak
energy condition.

However, in view of the fact that the cosmological term
can be regarded as a perfect fluid, we can also switch the 4D description
into a fluid with the following components,
\begin{equation}
8\pi {{T^{(4)}}^0}_0  = - 8\pi \rho = \alpha^2,
                      \label{eq:29}
\end{equation}
\begin{equation}
8\pi {{T^{(4)}}^r}_r  = 8\pi p = \alpha^2,
                        \label{eq:30}
\end{equation}
\begin{equation}
8\pi {{T^{(4)}}^\varphi}_\varphi  = 8\pi p = \alpha^2,
                        \label{eq:31}
\end{equation}
\begin{equation}
8\pi {{T^{(4)}}^z}_z  = 8\pi p_{zz} = 3 \alpha^2.
                        \label{eq:32}
\end{equation}
As in the 3D case, in this description, the cosmological term in
Einstein's equations (\ref{eq:4}), does not appear explicitly.
Metric (\ref{eq:24}) is thus a solution of Einstein's field equations,
${G^{(4)}}_{ab} = 8\pi  {T^{(4)}}_{ab}$, with
${T^{(4)}}_{ab}$ given in equations (\ref{eq:29})-(\ref{eq:32}).
Again, this fluid obeys the strong but not the weak energy condition.
It is a fluid with anisotropic pressures.

Thus,  the BTZ black hole can be translated into a black string
(or cylindrical black hole) in 4D general relativity. All the
features appearing in 3D, such as 3D collapse \cite{mannross} and 3D stars
\cite{zanelli} can be translated into 4D cylindrical general relativity.
The asymptotic symmetry group is $R$ times
the conformal group in two dimensions
of which $R\times SO(2,2)$ is a subgroup.
\vskip 1cm

\noindent
{\bf 5. Conclusions}

\vskip 0.3cm

It is remarkable that one can connect solutions in
theories with different dimensions. Here, we have connected 3D and
4D general relativity by an apropriate choice of metric and
energy-momentum tensor in 4D. Such a type of connection can also be
found from other 3D theories to 4D general relativity
\cite{lemos1,lemoszanchin}
as well as from 2D theories to 4D general relativity
\cite{call,lemos2}.

\end{document}